**Scaffolding fundamentals and recent advances in sustainable scaffolding techniques for cultured meat development**

AMM Nurul Alam[1], Chan-Jin Kim[1], So-Hee Kim[1], Swati Kumari[1], Eun-Yeong Lee[1], Young-Hwa Hwang[2], Seon-Tea Joo[1,2*]

[1] Division of Applied Life Science (BK21 Four), Gyeongsang National University, Jinju 52852, Korea

[2] Institute of Agriculture & Life Science, Gyeongsang National University, Jinju 52852, Korea

**Contact information for the Corresponding author**

Mailing Address: Institute of Agriculture & Life Science, Gyeongsang National University, Jinju 52852, Korea. E-mail stjoo@gnu.ac.kr

**Short version of title (running head):** Scaffolding for Cultured meat

**Abstract**

In cultured meat (CM) products the paramount significance lies in the fundamental attributes like texture and sensory of the processed end product. To cater to the tactile and gustatory preferences of real meat, the product needs to be designed to incorporate its texture and sensory attributes. Presently CM products are mainly grounded products like sausage, nugget, frankfurter, burger patty, surimi, and steak with less sophistication and need to mimic real meat to grapple with the traditional meat market. The existence of fibrous microstructure in connective and muscle tissues has attracted considerable interest in the realm of tissue engineering. Scaffolding plays an important role in CM production by aiding cell adhesion, growth, differentiation, and alignment. A wide array of scaffolding technologies has been developed for implementation in the realm of biomedical research. In recent years researchers also focus on edible scaffolding to ease the process of CM. However, it is imperative to implement cutting edge technologies like 3D scaffolds, 3D printing, electrospun nanofibers in order to advance the creation of sustainable and edible scaffolding methods in CM production, with the ultimate goal of replicating the sensory and nutritional attributes to mimic real meat cut. This review discusses recent advances in scaffolding techniques and biomaterials related to structured CM production and required advances to create muscle fiber structures to mimic real meat.

1. Introduction

The traditional meat production system is well recognized as a significant contributor to environmental deterioration. Livestock farming is responsible for close to fifteen percent of worldwide greenhouse gas emissions (GHG) and utilizes close to eighty percent of the habitable zone of earth for raising and nourishing cattle, moreover it only accounts for approximately 17% of the global calorie supply (Bonny et al., 2015; Swartz, 2019). The projected growth of the worldwide population by the year 2050, along with the rising demand for meat, limited availability of arable land, and constraints on animal production capacity, present a significant challenge to global food security (Aiking, 2011; Ray et al., 2013). Furthermore, the generation of waste in animal agriculture is a significant contributor to environmental pollution, primarily attributable to the consumer's inclination towards meat products. In the poultry sector, for instance, approximately 30% of the entire chicken carcass is often abandoned, with a bigger proportion being wasted in the bovine feed manufacturing business for protein meal (Welin, 2013). According to the Food and Agriculture Organization his contributes to the substantial quantity of food waste generated on a yearly basis (Gustafsson et al., 2013). It is always advisable to go for alternative meat production solutions like CM to deal with the issues related to meat production through livestock farming. CM holds the promise of creating meat that closely mimics the structure of traditionally farmed meat while minimizing the land required for production. The unsustainable practices involved in the existing techniques of meat production have necessitated the development of sustainable systems to create alternative protein sources in order to adequately nourish our future population (Datar & Betti, 2010).

CM is a nascent area of study that encompasses many technologies aimed at generating agricultural products that are biologically analogous to traditional meat through the cultivation of cell cultures. Utilizing stem cells CM production could be a viable and attractive substitute to conventional livestock-based meat production. Currently, researchers are studying simpler and new techniques to decrease the production cost of CM. A streamlined and enhanced technique utilizing a 10% concentration of equine serum has facilitated more effective cultivation of meat in comparison to prior methods. This breakthrough is anticipated to contribute to the creation of affordable and secure CM (Yun et al., 2023). According to, another study the simplest, cost-effective and time-efficient manner for purifying chicken satellite cells for CM is to pre-plate them at a temperature of 41°C for an uninterrupted duration of 4 hours (Kim et al., 2022). This advancement is anticipated to contribute to the creation of affordable and secure CM. Furthermore, meticulous monitoring of the production process inside a laboratory setting has the potential to eradicate foodborne infections like salmonella infection or mad cow disease. Moreover, the production of CM in conjunction with plant-derived sources has the potential to contribute to the economic aspect by mitigating costs and price, hence facilitating the growth of consumer demand. Notwithstanding these benefits, the process of introducing CM commercially to the market is impeded by numerous obstacles. It is imperative to note that the current CM falls short of satisfying consumer expectations in terms of both quality and quantity. Current CM products lack resemblance to real meat in terms of its structure, texture, color, flavor, sensory and nutritional composition (Fraeye et al., 2020; Pajčin et al., 2022; Zhang et al., 2021).

In order to precisely recreate the sensory and nutritional attributes associated with real meat, it is necessary to reproduce muscle development in a laboratory setting. Tissue engineering techniques can generate items that closely resemble real meat by introducing cells into an environment that reproduces natural tissue conditions (Bomkamp et al., 2022; Post, 2014). Prior and majority of current research efforts concentrated on the manufacturing process of CM (Benjaminson et al., 2002; Hanga et al., 2021; MacQueen et al., 2019; Singh et al., 2022; Stout et al., 2020). Scaffolds are porous structures that acts as templates for tissue formation (O'brien, 2011). They typically imitate the extracellular matrix, facilitating cellular attachment, proliferation, and/or differentiation. In CM, it is desired for the scaffold to assist in achieving certain desirable sensory features. Additionally, a crucial factor to consider is whether the scaffold should be an integral component of the end result, necessitating its edibility, compliance with food-grade standards, and nutritional value, or if it should be engineered to be easily removed. The primary goal of the scaffold is to promote the development of critical muscles, adipose tissue, and connective tissue. The final products derived from CM can undergo downstream processing to modify their properties, including texture and taste. These processing methods used are similar to those already used in the manufacturing of traditional meat products, such as chicken patties, minced beef, or sausages (King, 2019). In order to produce a structured CM product, such as a special cut, steak or fillet, it is necessary to employ a scaffolding approach that enables cell proliferation, differentiation into key cell types, and spatial organization.

Scaffolding is pivotal for achieving the desired appearance and texture of meat cuts, ultimately enhancing the consumer experience (Ong et al., 2020). However, in the case of meat tissue that is highly structured and organized, such as muscle cuts and steaks, the use of highly vascularized scaffolds would be necessary. Aiming to generate organized meat at a large scale, relying exclusively on cell proliferation within the bioreactor may be inadequate. There is a need for the adoption of three-dimensional scaffolding aided cultivation system that can efficiently support the growth and differentiation of cells, and perhaps enable self-assembly to achieve certain organoleptic characteristics. Integrated scaffolding techniques like tissue perfusion bioreactors have been the subject of research in the field of bone tissue engineering (Bhaskar et al., 2018; Gomes et al., 2006) and arterial applications (Williams & Wick, 2004), suggesting their potential suitability for cardiac muscle applications. While the utilization of scaffold materials and 3D printing technology has made it possible for the manufacturing of muscle cells that are closer to meat, there is still a limitation in terms of reducing production costs to a level where they can be utilized as food. Furthermore, the task of creating edible materials poses a barrier due to the present unsuitability of the materials employed for in the production of ingestible CM (D. Y. Lee et al., 2023). The primary aim of this review is to analyze the scaffolds used in cellular agriculture, as well as explore potential processing techniques and biomaterials that could be incorporated into these scaffolds to improve the structural qualities of CM, with the goal of mimicking the characteristics of real meat.

## 2. Cultured meat

Edible meat, denoting the musculature of farmed animals, predominantly consists of skeletal muscular components comprised of aggregations of muscular fibers. During muscle tissue formation, individual muscle cells known as myoblasts undergo fusion with one another, resulting in the formation of multinucleated myotubes. These myotubes subsequently unite to become muscle fibers (Ostrovidov et al., 2014). On the contrary, CM is produced through tissue engineering techniques using stem cells under controlled environment setting as illustrated in figure 1. Initially, a biopsy is carried out on the animal to obtain the initial cells. The cells endure proliferation and differentiation subsequently into skeletal muscle cells. These cells are then maintained till they reach their particular maturation stage. At this moment, they can be collected and processed to make the final product (Gaydhane et al., 2018). Collection of stem cells can be done from either live animal biopsies or freshly slaughtered animals and grown in vitro to multiply cells (Arshad et al., 2017; Ben-Arye & Levenberg, 2019). In CM generally satellite cells are utilized which are the adult stem cells found in skeletal muscles and have the ability to develop into skeletal myotubes with minimal external input. Additionally, these cells can be used to differentiate into either muscle or fat cells by isolating specific type of stem cell that is isolated (Post, 2014).

Various categories of stem cells are available for both research and biomedical use. A typical instance is stem cells coming from adipose tissue, which have the ability to differentiate into multiple kinds of cells, such as muscle, bone, cartilage, and fat cells (Datar & Betti, 2010). Induced pluripotent stem cells (iPSCs) were first described by Takahashi and Yamanaka (2006). iPSCs can be obtained through noninvasive methods without causing harm to animals. Another kind, embryonic stem cells (ESCs) are also commonly used in stem cell research (Amit et al., 2000) and since discovered till now under investigation in the field of CM (Kim et al., 2022).

The development of CM products, such as patties, meatballs, and chicken nuggets, has been extensively investigated. Research papers have provided strong evidence in favor of the feasibility of CM products (Cassiday, 2018). Muscle stem cells possess a high propensity for differentiating into myotubes and myofibrils, represent a compelling avenue for the development of CM. These cells are widely employed in this context (Asakura et al., 2001). In order to gain a substantial portion of the market for real meat, it is imperative to thoroughly examine and develop a broad range of products using CM. There is a need to create a CM production process that can enhance the various flavor attributes, such as umami and bitterness, and regulate the chemical make-up of the growing medium in order to generate CM that closely resembles the taste of actual meat (Joo et al., 2022).

### 3. Necessity of Scaffolds

In CM production stems cells placed on a biomaterial base known as a scaffold, microcarrier, or film. Scaffolds are kind of networks to provide support for the substrate, assist in the transportation of nutrients, and allow for cell respiration (Schätzlein & Blaeser, 2022). Scaffolds typically aids the replication of the extracellular matrix (ECM), providing a substrate for cell adhesion, proliferation, and differentiation. In CM process this scaffold base invigorates satellite cells to proliferate to build up a mass. Moreover, they enable the development of spatial heterogeneity within the resulting product, so achieving a meat-like structure (Stephens et al., 2018). Further development of edible directional scaffolds may replace the serum-free culture medium will significantly aid in reducing the CM production cost and animal welfare issues (Y. Chen et al., 2023).

In muscle stem cell growth, alignment, and adhesion are highly sustained by scaffold biomaterials. Scaffolding is essential for making dense, cohesive saturated meat to mimics flesh tissue's 3D milieu, particularly media perfusion and vascularization (Seah et al., 2022).

The primary objective of a scaffold is to replicate essential characteristics of the extracellular matrix (ECM), such as its mechanical integrity, flexibility, and nutritional composition. It is possible to replace the intricacy of the extracellular matrix (ECM) with less intricate scaffolding constructs that incorporate one or more essential structural proteins, growth factors, transcription factors, and other relevant components (Aisenbrey & Murphy, 2020).

## 4. Ideal characteristics of scaffolds

The scaffolds utilized in CM production must possess certain characteristics in order to effectively support tissue maturation. These characteristics include being biologically active, having a large surface area, being flexible and allowing for maximum growth medium diffusion. Additionally, it is important for these scaffolds to be edible, non-toxic, allergen free (Datar & Betti, 2010). The optimal porosity range for scaffolds is 30%–90%, with a pore size from 50 to 150 μm or higher for myogenic cells. The thickness of the scaffold usually depends on the processing techniques (Bomkamp et al., 2022). Microcarrier scaffolds are common in biomedical fields and offer multiple advantages compared to larger scaffolds. It has been observe that high cell densities and productivity are facilitated by their elevated surface area to mass ratio (Kong, Jing, et al., 2022). Microcarriers of bigger size enable improved cell attachment, whereas smaller microcarriers lead to higher growth rates as a consequence of increased shear stress (Norris et al., 2022).

Apart from microcarriers several other technologies are in use to manufacture scaffolds such as hydrogels, decellularized plants, 3D bioprinting, electrospinning, mold systems, injectable systems, etc. (Lee et al., 2019). While animal-derived extracellular matrix (ECM) proteins may not be suitable for use in cellular manufacturing (CM), it is possible to create these components using other methods such as microbial fermentation and plant molecular farming (Mohammadinejad et al., 2019). All the above techniques have some limitations such as unregulated dimension, inadequate consistency, residual chemicals, challenges in incorporating vascular networks and insufficient interconnectedness (Peng et al., 2018), which opens the door for further research to explore more suitable technologies and biomaterials in scaffold manufacturing.

According to previous studies using gelatin and soy protein aided in building ideal scaffolds to fulfill the necessary fibrous characteristics of meat (Ben-Arye et al., 2020; MacQueen et al., 2019). To mimic real meat the alignment of myoblasts is crucial and more investigation needs to develop aligned tissue structure. 3D scaffolds are routinely employed to facilitate cellular alignment and differentiation (Xiang et al., 2022) especially in the Biomedical fields. Technological challenges in scaffolding in CM seem highly complex due to the lack of a vascular system. To mimic real meat it is necessary to create a complex branching network of interrelated, consumable, versatile, and permeable material that allows for the circulation of nutrients and the attachment of myoblasts and other cell types (Bhat et al., 2014) which may be achievable by using appropriate scaffolds.

## 5. Muscle Tissue Structure

In the field of culinary arts, it is of utmost significance to comprehend the correlation between the structure of muscles and the nutritional as well as sensory aspects of meat. The development techniques of CM are closely connected to the muscle tissue structure as muscle shape links to the nutritional and organoleptic characteristics of meat. In current practices, satellite cells are mostly collected from the skeletal muscles of cattle, chicken, pork, lamb, and fish (Post et al., 2020). The complex hierarchical tissue structure is responsible for the fibrous structure of meat. The muscle fiber is a primary functional unit and is surrounded by intramuscular fat tissues, connective tissue, vasculature, nerve tissues and are the primary determinants of muscle texture and quality parameters (Bomkamp et al., 2022; Listrat et al.,

2016). To mimic the nutritional and structural attributes of real meat, CM techniques should be focused on recreating a tissue substantially composed of the same muscle structure like real meat muscle. Therefore, in order to devise pragmatic methodologies for the engineering of muscle tissue, it is imperative to understand the naturally existing muscular tissue (Figure 2).

The mouthfeel of intact meat pieces is attributed to a structured arrangement of proteins, organized in a certain order to produce elongated insoluble chains. These chains align in a parallel manner, ultimately forming elongated fibrils known as "myofibrils." The myofibrils further form the muscle fiber and are organized into elongated structures known as fascicles, which are the fundamental units of muscle tissue (Listrat et al., 2016). Myofibrils resemble like cables, which consist of contractile filaments made of elongated chains of actin and myosin. The filaments are partitioned into discrete contractile units known as sarcomeres. The distinct striated appearance of muscle cells is attributed to the arrangement of actin and myosin filaments that overlap within the myofibrils (Listrat et al., 2016; Lodish, 2008).

Intramuscular lipids play a crucial role in determining sensory and nutritional value of meat (Listrat et al., 2016) and plays a significant role in nutrition due to its provision of vitamins A, D, K, and E, as well as polyunsaturated fatty acids (Fish et al., 2020). Intramuscular fat predominantly consists of adipocytes, which are situated between muscle fibers and fascicles. The composition of intramuscular fat encompasses structural lipids, phospholipids, and intracellular lipid droplets located inside the muscle fibers (Listrat et al., 2016).

The primary source of load-bearing capacity in muscles is derived from the dense extracellular matrix (ECM), rather than the muscle fibers themselves. This highlights the significance of a robust support framework for fully developed muscle cells (Gillies & Lieber, 2011). In order to accurately replicate the mechanical properties of the extracellular matrix (ECM), it is imperative for tissue engineering approaches to employ scaffolding materials that closely resemble the mechanical characteristics of the ECM. This is crucial for achieving a faithful representation of the native tissue. Extracellular matrix (ECM) exerts significant influence on the quality of traditional meat, manifesting in both its biological influence on muscle fibers in living organisms and alterations occurring during the postmortem aging process (Nishimura, 2015).

## 6. Scaffolding Biomaterials

Structural properties of scaffolds can be influenced by the utilization of scaffolding biomaterials. Scaffolding biomaterials habitually exhibit notable attributes such as exceptional biocompatibility, elevated porosity, and the capacity to reinstate the extracellular matrix (ECM) (Sharma et al., 2015). In the biomedical field, synthetic polymers are commonly employed as substitutes for the extracellular matrix (ECM) due to their desirable characteristics such as ease of processing, ability to decompose naturally, and minimal immune response (Chen & Liu, 2016). But A buildup of breakdown products could possibly modify the pH of the surrounding tissue, leading to unfavorable inflammatory reactions. That's why, Edible biomaterials, particularly those of natural origin, have attracted significant interest and are extensively employed in tissue regeneration in the biomedical fields due to their easy availability similarity in biological and chemical properties to real tissue (Su et al., 2021). In order to go towards a more environmentally friendly and sustainable scaffold development for CM production, the biomaterials should incorporate elements that are renewable, cost-effective, and easily abundant.

Scaffolds so far utilized in cellular agriculture offered in several formats, including fiber/ single dimension scaffold, film/ two-dimension scaffold, and three-dimension scaffold (microcarrier, sponge, hydrogel, etc.). A wide variety of scaffolding materials like proteins, polysaccharides, and polynucleotides are used in CM production from animal, plant, or synthetic origin (Stevens et al., 2008). Mechanical stretching of myoblasts can be achieved through the utilization of porous materials composed of cellulose, alginate, chitosan, or collagen (Narayanan et al., 2020). Edible biomaterials like elastin, gelatin, collagen, and fibronectin, originating from animals are high in Extra Cellular Matrix (ECM) and stimulate greater cellular growth (Reddy et al., 2021). Researchers are continuously studying different options and combinations of materials for producing sustainable scaffolds such as plant protein, animal protein, fish protein, marine animals, fungi, bread, seaweed, fruits, decellularized plants etc. A scaffold was recently created via the combination of gelatin and soymilk on which C2C12 cells exhibited a high expression of myosin, a crucial protein for myotube formation (Li et al., 2022). Different conventional and unconventional biomaterials so far used in the biomedical cell culture and CM has been illustrated in Table 1 for a better understanding of their efficacy and guideline for future research on scaffolding techniques.

To explore potential of different biomaterials as a scaffold for CM production requires additional investigation and development. Significant developments in the development of scaffolds discussed in detailed on the following sections.

## 7. Edible scaffolds

For CM production edibility is the necessary need for scaffolds to ensure they possess the intended taste, nutrient composition, and texture (Norris et al., 2022). Thus, exploring edible and food-grade scaffolding materials is necessary to produce more sustainable and safer CM. There is a demand for an ingestible and esculent framework that is suitable for nurturing the growth of animal muscle tissue. Consumable material eases the process of CM by abolishing the step of separating scaffold from tissue (Enrione et al., 2017). Inedible and/or toxic compounds including solvents and crosslinkers should be minimized through intensive care to the different biomaterial processing steps. Researchers are deliberately working on the development of edible and biodegradable scaffolds for CM production (Post et al., 2020).

Different efforts of manufacturing edible scaffolds have been discussed here for better understanding of the pros and cons for future sustainable research and developments. An edible scaffolds was produced using a method of cold-casting using micro molding, in which The films have been created using non-mammalian components (Acevedo et al., 2018). Orellana et al. (2020) developed a composite preparation possessing unique characteristics like bioactivity (salmon gelatin), crosslinking (calcium alginate), gelling (agarose), and plasticizing (glycerol).

Although the incorporation of non-animal materials is encouraged for cell CM, the efficient and proper processing of animal by-products such skin, ligaments, intestine and bones may also contribute to improving the sustainability of the traditional livestock farming. The application of animal by-products such as collagen (Ferraro et al., 2016) can be effectively deployed as edible casings, hence mitigating environmental contamination. In lieu of utilizing animal by products, a scaffold including aligned porous structures was developed using the directional freeze-drying technology applied to an edible collagen hydrogel which offered porous space for cell adhesion and growth and enhanced the directional development of cells due to plentiful tripeptide (Y. Chen et al., 2023).

Plant proteins and polysaccharides from different sources are frequently preferred as biomaterials for cell culture meat scaffolds. Decellularized spinach leaves as an edible scaffold were found to be similar to gelatin-coated glass to develop CM and was cost effective, potential, sustainable and ecological (Jones et al., 2021). A List of decellularized scaffold material options given in table 1 for better understanding and future research directions. Plant proteins such as soy, pea, zein and glutenin are competitive, abundant and able to transformed into films with appropriate mechanical characteristics for CM development (Dong et al., 2004). The protein films made from glutenin and zein effectively promoted the growth of aligned cells and the further development of aligned myotubes (Xiang et al., 2022). Glutenin and zein could be promising candidates for upcoming research in the production of CM. Textured vegetable proteins (TVP) can be used for making fibrous scaffolds (Bakhsh et al., 2022). In addition to lowering the price of scaffold and CM, it could be beneficial to look for additional sources of premium plant protein that are available at affordable costs.

## 8. Electrospinning

Electrospinning is a highly adaptable and efficient method that is utilized in several disciplines, including textiles, structural engineering, nano-devices, filtration, health safety items, biological fields, and tissue engineering. It offers a user-friendly interface, cost-effective with little solution utilization, Easy to replicate and modify with adjustable fiber size and the ability to be tailored to specific requirements (Thenmozhi et al., 2017).

The manufacture of nano fibers through the application of electrostatic forces has been recognized and examined by scientific and technological studies for a period of 120 years. In 1887, Charles Vernon was the first to address the manufacturing of nanofibers. He described the process of obtaining thin glass fibers, likely with a diameter smaller than a micrometer, employing a technique called "drawing" (Tong & Sumetsky, 2011). Nevertheless, the exploration of electrostatic forces began much earlier with Gilbert during 1600, who noticed the formation of a cone and jet when a charged amber stone was brought close to a droplet (Luo et al., 2012). John Francis Cooley submitted the first patent for electrospinning in 1900, marking a major breakthrough for future study (Agarwal et al., 2016). In 1914, researchers assessed the behavior of a droplet produced on a metallic capillary tip, which enabled them to mathematically simulate fluids that are exposed to electrostatic forces (Rafiei & Haghi, 2015).

Carbohydrates, such as monosaccharides, oligosaccharides, and polysaccharides, have practical applications in the food delivery industry through electrospinning techniques. These carbohydrates can be classified based on their biological sources, which include higher plants (such as cellulose, starch, pectin), animals (such as chitosan, chitin), algae (such as alginate and carrageenan), and microorganisms (such as xanthan, dextran, and cyclodextrins) (Fathi et al., 2014). Cellulose, Chitosan and their derivatives starch and dextran have been identified as suitable materials for electrospinning outer wall materials (Desai et al., 2008; Ghorani & Tucker, 2015; Kong & Ziegler, 2012; Sun et al., 2013). Researchers created edible nanofibrous thin films using electrospinning (Wongsasulak et al., 2010).

The alignment and the porous architecture within the scaffolds known to serve a significant role in emulating the structural characteristics of real meat. In CM production it is imperative to explore innovative approaches for fabricating aligned scaffolding architectures like electrospun nano fibers. Even though electrospinning in the food processing is largely employed in food packaging and delivery systems (Coelho et al., 2021; Khoshnoudi-Nia et al., 2020), the possibility of this technique to generate microfibers has piqued the interest of researchers for CM technologies.

Electrospinning is a simple method that only requires a few mechanical component: high voltage power supply, syringe pump, syringe with a needle or capillary tip (spinneret), and collector (plate/drum) (Esfanjani & Jafari, 2016). The electrospinning technique requires the application of electric field with high-voltage to charge the surface of a droplet containing a polymer solution. This charge enables a liquid jet to be released through a spinneret (Mendes et al., 2017). Electrospinning can be divided into two distinct phases, at first, a biomaterial solution is loaded into a syringe and injected through the needle with the high voltage electric field, where spherical droplets form into a Taylor cone (Figure 3). Later the solvent evaporates to get thinner and conical droplets solidly to create micro or nanofiber filaments which deposits on a grounded or rotating collector (Ghorani & Tucker, 2015). The electrospinning technique is known for its efficient, low-cost, and sustainable process to produce fine fibers using biopolymer materials (Zhang et al., 2018). Electrospinning is intriguing for scaffold construction because it is easy, controlled, have extensive surface area (Rijal et al., 2017), reproducible, and it can produce polymers with distinct properties (Riboldi et al., 2005).

Nanofiber scaffolds should be composed of safe and edible components like polysaccharides and proteins that are well suited for CM production because they can easily break down, are compatible with living cells, have biological effects, and can be safely consumed (Ben-Arye & Levenberg, 2019). Most of the research works on nano fibers using electrospinning technique are in the biomedical field. Biomedical studies conducted to evaluate the influence of permeability and orientation of random and aligned electrospun nanofibers on skeletal muscle cells (Park & Lek, 2016). Immersion rotary jet spinning technique used to focus CM and scaffold constructed with gelatin which was seeded with bovine aortic smooth muscle cells and rabbit skeletal muscle myoblast to mimic muscle tissues (MacQueen et al., 2019). This study was not successful to produce a final product to mimic natural contractile muscle. There have been limited research utilizing electrospinning techniques to create nano fibers for the purpose of cell proliferation or adhesion in the CM production process. However, this method is gaining popularity since it aims to replicate actual flesh. Figure 3 provides a concise overview of several current research papers, offering a deeper understanding of electrospinning techniques in the CM field.

Alternative spinning processes, like as rotary jet spinning, can be employed to generate nanofibers possessing diverse advantageous characteristics for CM (MacQueen et al., 2019). Needleless electrospinning has developed as a more effective method for large-scale manufacture of ultrathin fibers as opposed to conventional needle electrospinning (Yu et al., 2017). In needleless electrospinning the maintenance is easier and no clogging of spinneret found than needle e-spinning. Whereas it is difficult to maintain the consistency of the solution concentration and viscosity (Yu et al., 2017). In theory, the e-spinning jets can be generated from any type of spinnerets, provided that the electrostatic force surpasses the critical limit of the electrospinning solution (Niu et al., 2012). The same needle-free approach can be employed to rotary spinnerets that are partially submerged in the polymer solution (Jirsak et al., 2010). Moreover, needleless approach may be useful to avoid the clogging of needle for many non-animal based biomaterials in the CM production. Nevertheless, this type of techniques needs to be studied and verified for the possible development of nanofiber scaffolds for CM production.

### 9. Three Dimensional (3D) bioprinting

The utilization of three-dimensional (3D) printing in meat processing is already underway, since it offers the added advantages of producing bespoke geometrics and large-size structures (Dick et al., 2019). 3D bioprinting is a specialized area within 3D printing that focuses on the precise organization of bioinks, which are composed of a combination of biomaterials and living cells, in a step-by-step manner and enables the fabrication of three-dimensional biological structures (Murphy & Atala, 2014). Tissue engineering (Chu et al., 2021; Demirtaş et al., 2017), drug screening (Ma et al., 2018; Nie et al., 2020), and regenerative medicine (S. Y. Lee et al., 2023) are only couple of areas that have made substantial use of this technology due to its remarkable efficacy, precision, and accuracy.

The application of three-dimensional bioprinting into the production procedure of CM is now in its preliminary phases. The current state of 3D printed meat is still dependent on components that are derived from animals. According to S. Y. Lee et al. (2023), 3D bioprinting has brought about major advancements in the fabrication of CM, which is a field of research that has been deemed to be at the cutting edge. Tissue engineering and 3D printing combinedly has the potential to significantly enhance the quality of CM by the creation of a product that closely resembles the appearance of animal flesh (Balasubramanian et al., 2021; K. Handral et al., 2022; Lanzoni et al., 2022). Previous research has provided theoretical concepts for this methodology and theorized about the potential advantages that it may give. In spite of this, these works do not contain a in depth review of the progress that has been made in this particular field, and there has been no special evaluation that has been conducted to offer direction for the future progression of this research direction.

3D bioprinting focuses on the concerns in the existing scaffolding techniques utilized in CM and trying to effectively govern tissue maturation, potentially leading to suboptimal nutritional and organoleptic characteristics in CM. The utilization of 3D bioprinting techniques enables exact control over the ratios of cells, their spatial arrangement, and the densities of specific cell types (Gungor-Ozkerim et al., 2018). In the recent years, there has been an increasing interest among academics in exploring various bioink formulations for their usefulness in the field of tissue engineering and regenerative medicine. This interest has been spurred on by the advent of 3D printing. A summary of the most important discoveries and developments in bioink research during the past few years may be found in Table 2.

This advancement in bioprinting holds the potential to produce CM that closely resembles traditional meat in terms of texture and overall characteristics. Fabrication of 3D tissue construction is possible by bio printing with pre-programmed structures and geometries containing biomaterials and/or living cells, this combination is known as bioink (Gungor-Ozkerim et al., 2018). The process of 3D bioprinting distinguishes itself from the exclusive cultivation of meat due to its ability to fabricate muscle cells, fat cells, and extracellular matrix

(ECM) supporting cells within a scaffold that facilitates cellular growth and proliferation (Sun et al., 2018). Following the completion of the printing procedure, the meat undergoes an additional maturation phase, typically facilitated by bioreactors that facilitate the transportation of nutrients (Zhang et al., 2018).

The fibrous composition of a 3D printed scaffold plays a crucial role in augmenting the sensory attributes of printed meat products. 3D bioprinting was successfully used to mimic porcine skeletal muscle tissue. Ozbolat & Hospodiuk suggested that extrusion-based bio printing is feasible for constructing CM scaffolds. Furthermore, they can imitate natural tissues which is convenient (Ozbolat & Hospodiuk, 2016). A combination of pea protein isolate and soy protein isolate with RGD-modified alginate is recommended as potential bioinks for 3D printed scaffolds in bovine satellite cell cultivation with up to 90% cell viability (Ianovici et al., 2022). Using cellularized gel fibers, beef steak like tissue was created from bovine satellite cells and printed onto a support bath with tendon-gel integrated bioprinting-TIP technique (Kang et al., 2021).

Some researchers mention other methods than 3D bioprinting with biomaterials may have bottlenecks like food incompatibility, residual effects, high cost, etc. (Levi et al., 2022). The rheological characteristics of meat-based ink, the deposition process after printing, and the stability of the printed scaffold during post-processing (cooking) are crucial variables that must be considered in order to achieve the necessary requirements for 3D bioprinting of CM. For CM production, 3D bio printing technology may promise way to obtain tissue structure from muscle cells.

Companies started with meat alternative extended their investments for the commercial production of CM and fish. Figure 4 illustrates the major players in 3D CM production investors and their latest standpoint in the industry. Companies that initially entered the market of meat analogs have expanded their investments to 3D bioprinting facilities in large scale for the commercial production of cultured meat and fish. Figure 4 depicts the key stakeholders involved in the manufacture of three-dimensional (3D) bioprint generated manufacturing of CM, together with their most recent positions within the industry.

**10. Conclusion**

1. Intensive research is required to bring the organoleptic and structural characteristics of real meat into CM. In order to mimic real meat, the design or structure of scaffold would be of paramount important in coming days. 3D scaffold or nanofilms/fibers could be helpful for cell attachment in a structured way.
2. Wide range of biomaterials studied for creating scaffolds to culture animal cells, but sustainable biomaterials need to be inexpensive, nontoxic, edible, degradable and easily available. Many edible conventional and unconventional materials such as plant protein, decellularized leaves, plant polysaccharides, algae, insect, breads have been used as scaffolding materials and noted in this review which are natural, having significant nutritional value and safe. Further study on these biomaterials will be helpful to develop sustainable scaffolds.
3. Although different types of scaffolding techniques are in use for the production of CM, the major focus should be the development of structured product through aligned and directional cell growth to mimic the real meat muscle cut. With this focus, electrospinning techniques and 3D bioprinting could be promising for the creation of elongated nanofibers, directional cell growth and structured end product considering suitable edible biomaterials that are well adapted with these processes.

**Acknowledgment**: This research was supported by a National Research Foundation of Korea (NRF) grant funded by the Korean government (MSIT) (No. 2020R1I1A206937911), and the

Korean Institute of Planning and Evaluation for Technology in Food, Agriculture, Forestry and Fisheries (IPET) through the Agri-Bioindustry Technology Development Program, funded by the Ministry of Agriculture, Food, and Rural Affairs (MAFRA) (Project No. 321028-5), Korea.

**Conflicts of Interest:** The authors declare that they have no conflicts of interest.

**Table 1. Scaffolding materials in Muscle cell culture and Cultured meat.**

| Origin | Category | Scaffolding material | Species | Achievements | References |
|---|---|---|---|---|---|
| Plant | Protein | Textured Soy Protein (TSP)/ Textured Vegetable Protein (TVP), | Cattle | Alleviate bovine skeletal muscle cell growth and develop into a Cultured meat prototype. | (Ben-Arye et al., 2020) |
| | | | | Plant protein-based scaffolds seem appealing due to their inexpensive and additional nutritional content biocompatibility. | (Gutowska et al., 2001) (Radomsky et al., 1998) |
| | | Wheat | | Although wheat protein is a fascinating source of protein, its widespread use is limited due to allergen issues and | (Czaja-Bulsa & Bulsa, 2017) |
| | | | | inflated public impression of wheat protein intolerances. | (Palmieri et al., 2018) |
| | | Peanut | Pig | Scaffold derived from peanut protein aided flawless adhesion, the proliferation of pig smooth muscle cells with high cell survival rate and high-quality final product. | (Zheng et al., 2022) |
| | | Soy and Peanut | | Soy and peanut proteins are suggested to be labeled as allergens, although these are sustainable and low-cost. | (Post et al., 2020) |
| | | Zein and Protein Isolates from Soy, Wheat, Oat, Cottonseed, Pea, and Peanut | | Low cost, beneficial nutritional values, extensively used in the food industry, and well-known to consumers can make this a highly promising candidate as scaffold items in cultured meat manufacturing. Insufficient cell adherence might require protein modifications or incorporation of cell adherence moieties. | (Cai et al., 2020) |
| | Polysaccharide | Alginate, pectin, Konjac gum, Cellulose | | Potential biomaterials due to their physiological functions and substantial cellular adhesion compatibility. | (Levi et al., 2022) |
| | | Dextran | Cattle | bovine myoblast cells successfully cultured on microcarrier scaffolds prepared with dextran and polystyrene. | (Verbruggen et al., 2018) |
| Decellularized plant | Polysaccharide | Spinach | | Edible, cost-effective, and similar to gelatin-coated glass to develop Cultured meat. | (Jones et al., 2021) |
| | | Spinach | | Studied on human mesenchymal cell culture. | (Robbins et al., 2020) |
| | | Broccoli | | Broccoli florets were grown to show that they can support BSCs in a dynamic reactor. Decellularized florets had good physical and nutritional properties, suggesting their possible use in cultured meat production and consumption. | (Thyden et al., 2022) |
| | | Amenity grass leaf | Mouse | Cell alignment and proliferation of mouse C2C12 cells were assisted with natural morphology. | (Allan et al., 2021) |
| | | Green-Onion | Human | Studied on human mesenchymal cell culture. | (Cheng et al., 2020) |
| | | Apple, Carrot, and Celery | | Studied on mice C2C12 cell culture. | (Contessi Negrini et al., 2020) |
| | | Celery | | Studied on mice C2C12 cell culture. | (Campuzano et al., 2020) |
| Fruits | Polysaccharide/Protein | Jackfruit | | Artificial intelligence-based study showed that jackfruit scaffolds can grow pig myoblast cells. This method preserved the marbled cell appearance. The jackfruit scaffold also browns like meat when cooked. | (Ong et al., 2021) |
| Sea Weed | | Agar/agarose | Mouse | Utilized effectively in many investigations to implant the mouse myoblast cell line. | (Cidonio et al., 2019; Garcia-Cruz et al., 2021) |
| | | | | These polysaccharides can form stable and tasteless hydrogels but without any nutritional value. | (Delcour & Poutanen, 2013) |
| | | Marine macroalgae | Mouse | The decellularization-recellularization technique was employed to construct scaffolds made from seaweed cellulose for the purpose of culturing C2C12 cells in mice. | (Bar-Shai et al., 2021) |
| animal | Protein | Collagen | | Bio-engineered artificial muscles (BAM) found to grow successfully into collagen scaffolds and stand as the gold standard. | (Lanza et al., 2020; Leong et al., 2003; Liao et al., 2002; Lu et al., 2000; Taqvi & Roy, 2006; Yang et al., 2002) |
| | | Collagen | Chicken | The utilization of collagen microcarriers was investigated in order to enhance the process of myogenesis and replicate the characteristics of muscle fibers resembling natural tissue. | (Yang et al., 2022) |
| | | Gelatin | Cow and Rabbit | Gelatin, a natural component of meat generated via collagen denaturation during processing and cooking, has recently been employed to develop meat. | (MacQueen et al., 2019) |
| | | Gelatin | | The utilization of an edible gelatin microcarrier scaffold facilitated the development of myogenic microtissue from C2C12 or bovine satellite muscle cells. | (Norris et al., 2022) |
| | | Gelatin | Pig, mouse | Bovine gelatin-coated microbeads hydrogel has potential for primary porcine myoblast and C2C12 cell adhesion | (Kong, Ong, et al., 2022) |
| | | Fibrin | Cattle | Bovine myogenic cells were cultured on fibrin hydrogel to induce the development of a cellular arrangement that closely resembles the natural alignment observed in vivo. | (Takahashi et al., 2022) |
| | | Collagen, Fibrin | Cattle | Aligned bovine myotubes cultivated on collagen/fibrin hydrogel scaffold with low microbial level below the detection limit, indicating sterile tissue unlike commercial meat. | (Furuhashi et al., 2021) |
| | | Laminin, Gellan gum | Mouse | Laminin/gellan gum hydrogel scaffold facilitate mouse C2C12 skeletal muscle regeneration. | (Alheib et al., 2022) |
| | | Salmon gelatin | Mouse | Salmon gelatin was incorporated into the scaffolds and films to promote cell adhesion. Moreover, suitable myogenic responses were observed. | (Enrione et al., 2017; Orellana et al., 2020) |
| | | Fish gelatin | Mouse | Fish gelatin/agar matrix was optimized to coat the surface of textured vegetable protein (TVP), a plant-based cellular scaffold. | (Lee et al., 2022) |
| | | Fibrinogen | | Possess attractive nutritional values, cell adherence, and growth. | (Ng & Kurisawa, 2021) |
| | | Whey | | Low cost, sustainable but Need allergen labeling. | (Post et al., 2020) |
| Marine animals | Protein | Edible insects | | They are developing recognition due to their elevated protein and fat content, as well as their more nutritional and long-lasting value. | (Verbeke et al., 2015) |
| | | Chitosan | | It has the requisite cytocompatibility, is edible, inexpensive in cost, and widely available. Chitosan is a biodegradable biomaterial from marine source. However, its animal origin may limit its acceptance. | (Levi et al., 2022) |
| | | Chitosan, Collagen | Mouse, rabbit, sheep, cattle | The utilization of microcarriers composed of chitosan and collagen facilitated the adherence and swift proliferation of many cell types, including mouse skeletal C2C12 myoblasts, rabbit smooth muscle cells, sheep fibroblasts, and bovine umbilical cord mesenchymal stem cells. | (Zernov et al., 2022) |
| Fungi | Polysaccharide | Enoki mushroom | | Enoki mushroom is found to be edible polysaccharides and is promising as biomaterials to fabricate scaffolds. | (Hu et al., 2019; Wang et al., 2018; Zhang et al., 2012) |
| Improvised biomaterials | Protein | Sodium alginate and Gelatin gel coated with Tea polyphenols | | The adhesion rate of cells on scaffolds with a coating was 1.5 times more than uncoated, along with increased cell proliferation. The Cultured meat obtained from rabbits had comparable visual and sensory attributes to that of freshly harvested meat. | (X. Chen et al., 2023) |
| Bakery | Polysaccharide | Bread | | A compelling choice for future applications due to their ability to combine cost-effectiveness with simplicity, hence enabling scaled manufacturing. | (Holmes et al., 2022) |

**Table 2. Details on recent research on 3D bioprinting.**

| Bio Ink category | | | | Cell origin | Research focus | Reference |
|---|---|---|---|---|---|---|
| Chemical | Animal | Insect | Plant | | | |
| Alginate | Gelatin | | | 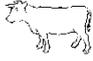 | Cultured meat | (Dutta et al., 2022) |
| RGD-modified alginate | | | Pea protein isolate (PPI), Soy protein isolate (SPI) | 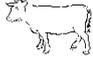 | Cultured meat | (Ianovici et al., 2022) |
| Alginate | Bovine adipose tissue | | | 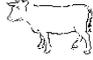 | Cultured meat | (Zagury et al., 2022) |
| | Fibrin, Gelatin | | | 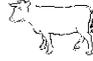 | Cultured meat | (Kang et al., 2021) |
| | Gelatin | | | 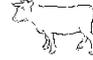 | Cultured meat | (Jeong et al., 2022) |
| | Gelatin | | Transglutaminase | 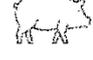 | Cultured meat | (Liu et al., 2022) |
| Alginate | Gelatin | | | 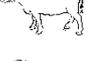 | Cultured meat | (Zanderigo, 2021) |
| | Gelatin | | | 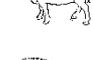 | Cultured meat | (Garrett et al., 2021) |
| Alginate | Gelatin | Silk/Fibroin | | 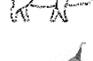 | Cultured meat | (Y. Li et al., 2021) |
| | | | Soybean oil-based resin | 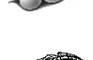 | Plant based meat | (Sealy et al., 2022) |
| Alginate | Gelatin | | | 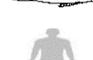 | Cultured fish | (Xu et al., 2023) |
| Alginate | Collagen, Gelatin, Fibrin | | Carrageenan | 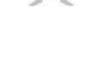 | Biomedical cell culture | (Machour et al., 2022) |
| Carrageenan | | | Carrageenan | 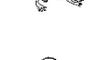 | Biomedical cell culture | (Marques et al., 2022) |
| Alginate | Gelatin | | | 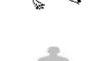 | Biomedical cell culture | (Bolívar-Monsalve et al., 2021) |
| | Fibrin, Gelatin | | | 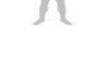 | Biomedical cell culture | (Ahrens et al., 2022) |

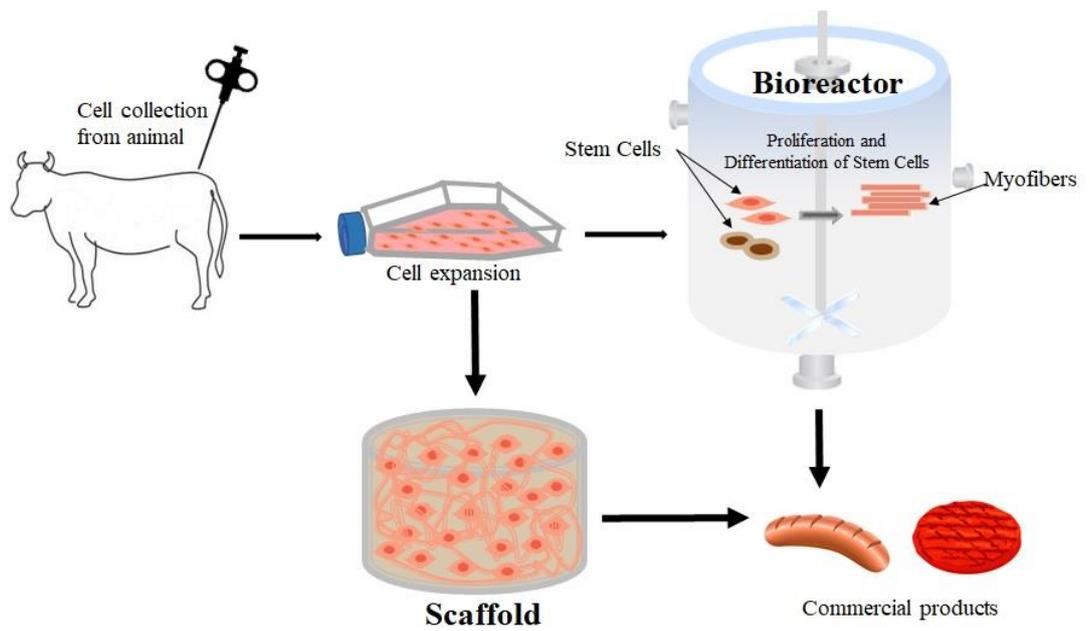

**Figure 1. illustration of lab grown meat (Cultured meat). Self-drawn images.**

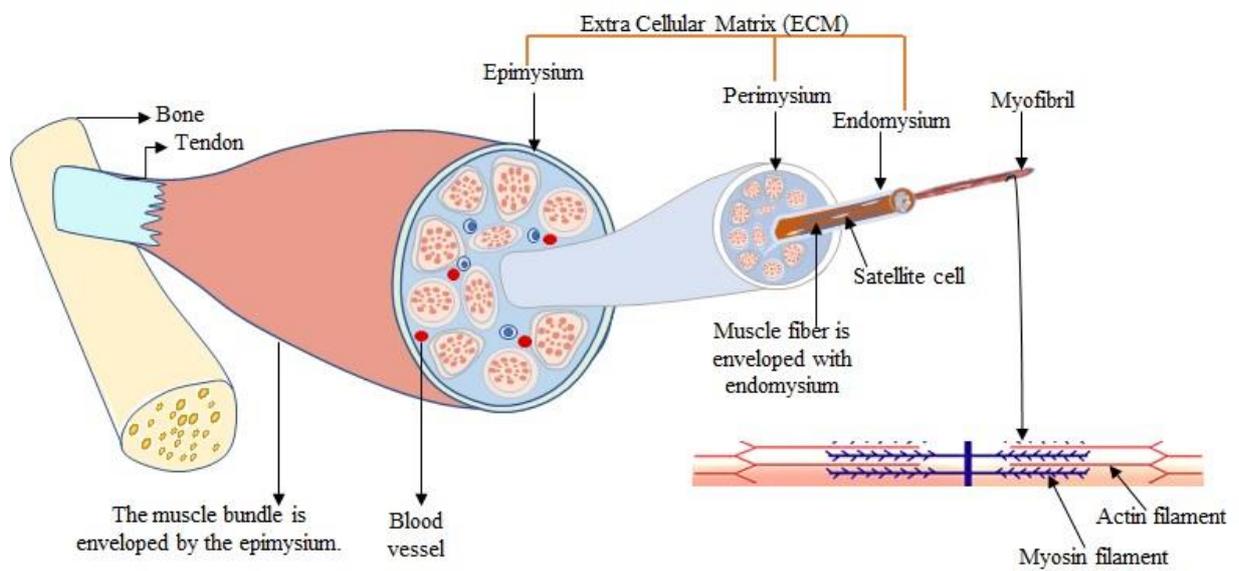

**Figure 2. skeletal muscle structure with extracellular matrix (ECM). Self-drawn images.**

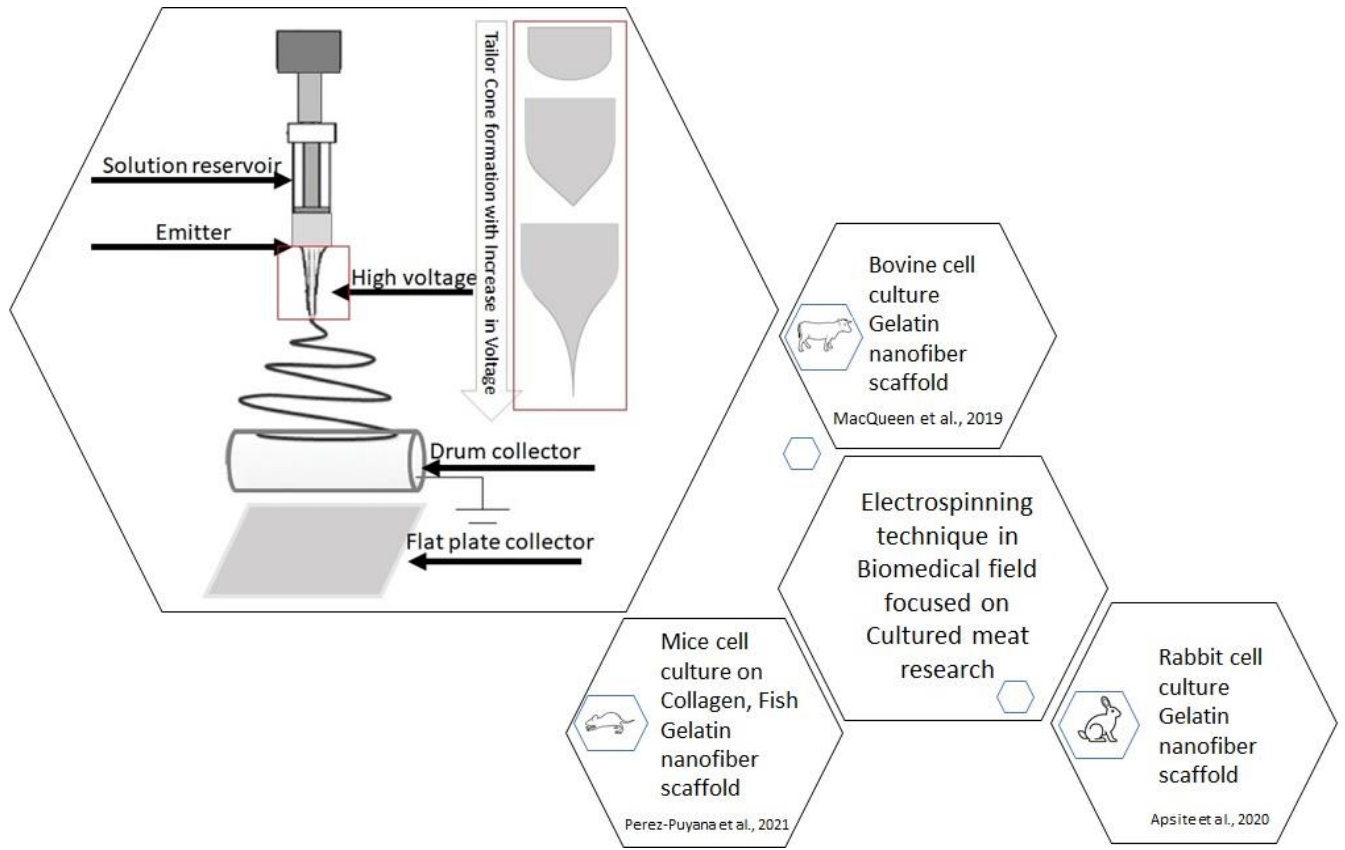

**Figure 3. Electrospinning process and recent uses in cell culture (Apsite et al., 2020; MacQueen et al., 2019; Perez-Puyana et al., 2021). Self-drawn images.**

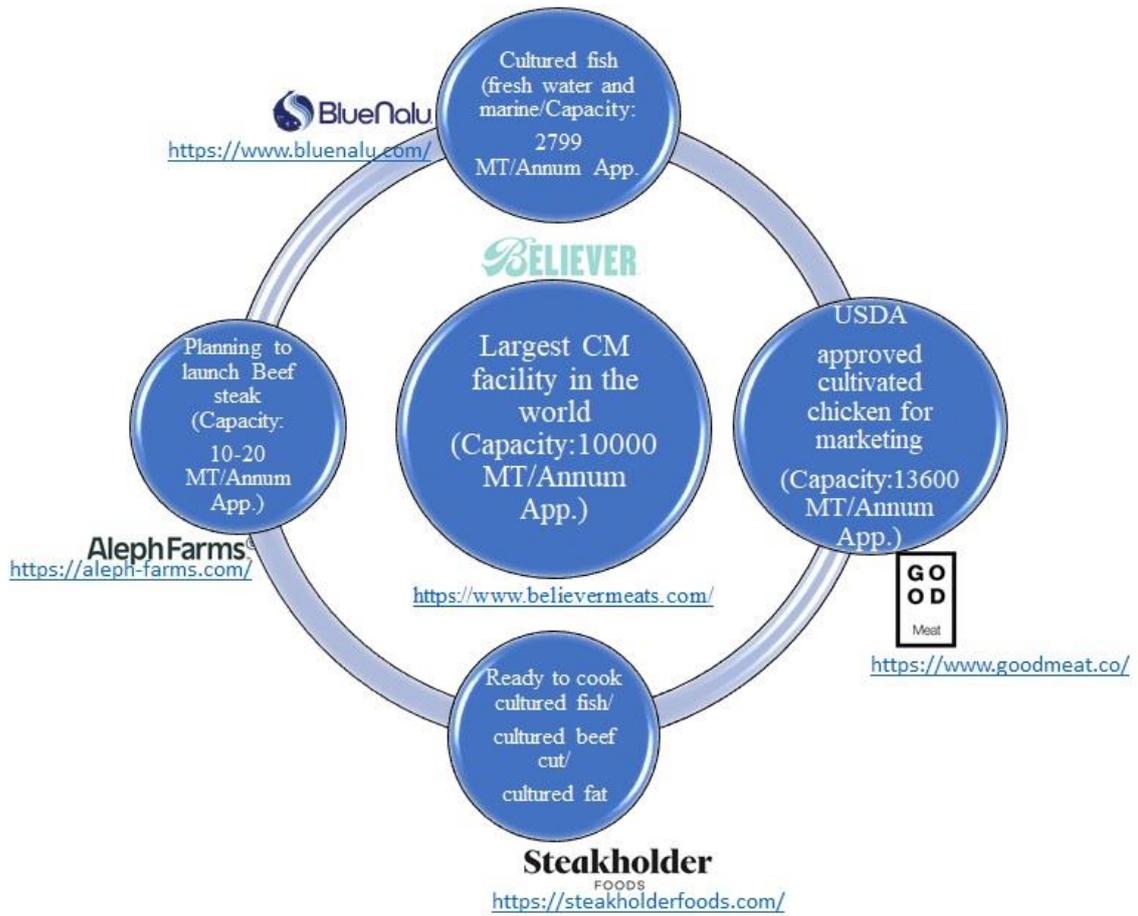

**Figure 4. Top companies in 3D Cultured meat development (Self-drawn images).**